\documentclass[preprintnumbers,amsmath,amssymb,twocolumn,nopacs,prb]{revtex4}
\usepackage{graphicx}% Include figure files
\usepackage{pifont}
\usepackage{dcolumn}% Align table columns on decimal point
\usepackage{bm}% bold math

\usepackage{amssymb}
\usepackage{amsmath}
\usepackage{array}

\begin{document}

\title{Optimizing synthetic diamond samples for quantum sensing technologies \\
by tuning the growth temperature}

\author{S.~Chouaieb$^{1}$, L.~J.~Mart\'{\i}nez$^{1,2}$, W.~Akhtar$^{1}$, I.~Robert-Philip$^{1}$, A.~Dr\'eau$^{1}$, O.~Brinza$^{3}$, J.~Achard$^{3}$, A.~Tallaire$^{3}$ and V.~Jacques$^{1}$}
\affiliation{$^{1}$Laboratoire Charles Coulomb, Universit\'{e} de Montpellier and CNRS, 34095 Montpellier, France}
\affiliation{$^{2}$Center for Quantum Optics and Quantum Information, Universidad Mayor, Camino La Pir\'{a}mide 5750, Huechuraba, Chile}
\affiliation{$^{3}$Laboratoire des Sciences des Proc\'ed\'es et des Mat\'eriaux, Universit\'{e} Paris 13, Sorbonne Paris Cit\'e and CNRS, 93430 Villetaneuse, France}

\begin{abstract}
Control of the crystalline orientation of nitrogen-vacancy (NV) defects in diamond is here demonstrated by tuning the temperature of chemical vapor deposition (CVD) growth on a (113)-oriented diamond substrate. We show that preferential alignment of NV defects along the [111] axis is significantly improved when the CVD growth temperature is decreased. This effect is then combined with temperature-dependent incorporation of NV defects during the CVD growth to obtain preferential alignment over dense ensembles of NV defects spatially localized in thin diamond layers. These results demonstrate that growth temperature can be exploited as an additional degree of freedom to engineer optimized diamond samples for quantum sensing applications.
\end{abstract}
\maketitle

The negatively-charged nitrogen-vacancy (NV) defect in diamond is currently one of the most promising solid-state quantum system for a broad range of emerging applications in quantum technologies~\cite{DefectReview}. Besides its use as a qubit for quantum information protocols~\cite{Dutt2007,Waldherr2014,Hensen2015,Kalb2017}, the electronic spin of the NV defect can be employed as an atomic-size quantum sensor providing an unprecedented combination of spatial resolution and sensitivity to several physical quantities including temperature~\cite{Neumann2013,Toyli2013,Kucsko2013}, electric~\cite{DoldeNatPhys} and magnetic fields~\cite{Maze,Gopi,Rondin_2014}. During the last decade, performance improvements of NV-based quantum technologies have been closely linked to the design of innovative diamond samples hosting NV defects with optimized properties. Depending on the target application, various parameters must be controlled such as NV density~\cite{Acosta2009}, spin coherence time~\cite{Balasu2009,Mizuochi}, charge state~\cite{Grotz2012,Pfender2017} and proximity from the sample surface~\cite{DegenPRB2012,TetiennePRB2018}. In this work, we focus on the control of the crystalline orientation of the NV defect, which defines its intrinsic spin quantization axis. \\
\begin{figure*}[t]
\includegraphics[width = 18.1cm]{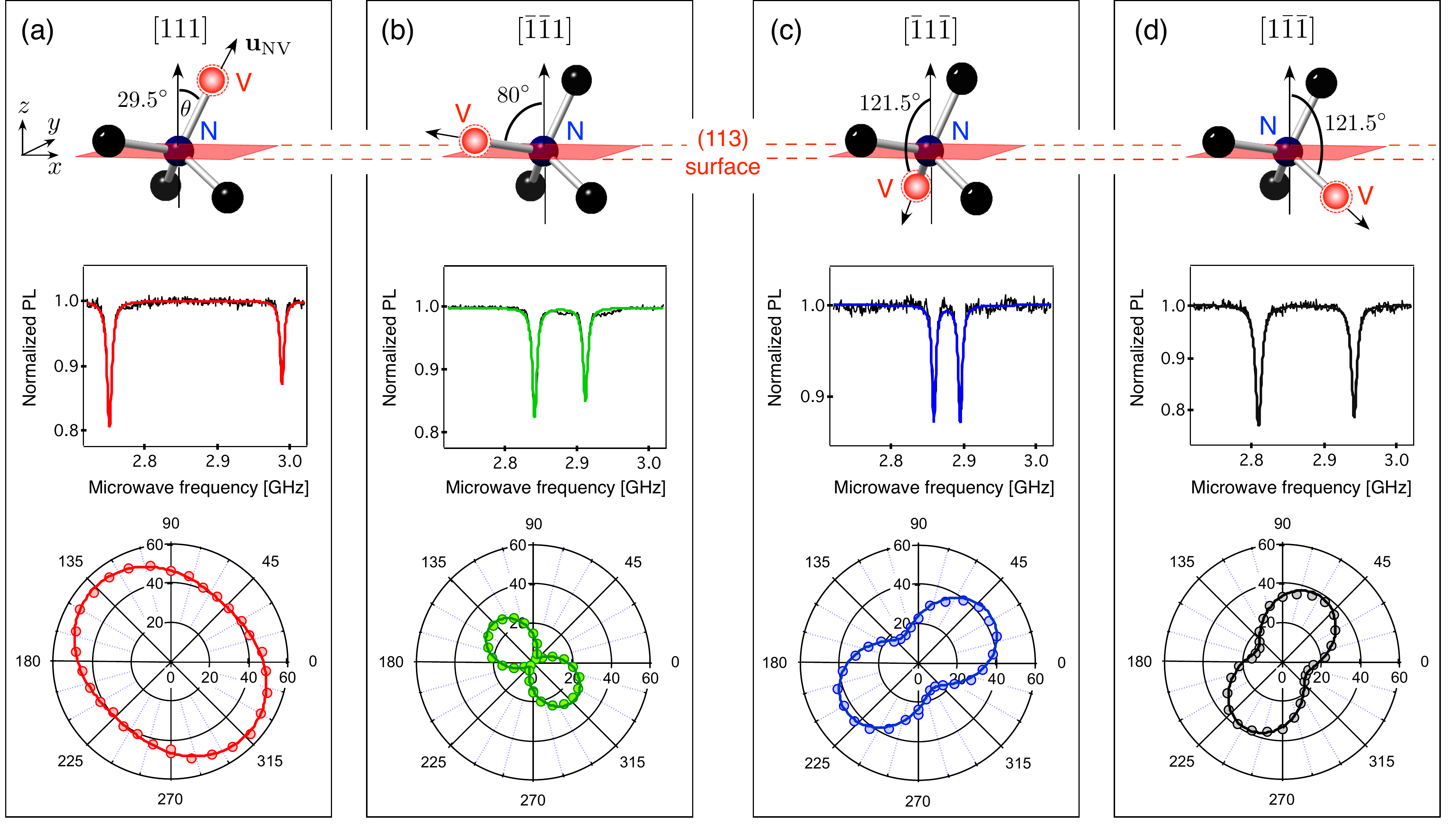}
\caption{(a) to (d) (top panels) Schematic drawing of the four possible NV defect orientations in a (113)-oriented diamond sample. (middle panels) Orientation-dependent ESR spectra recorded from single NV defects while applying a static magnetic field $\mathbf{B_0}$. (bottom panels) Corresponding polarization-dependent PL intensity diagrams. These measurements were performed at low laser power $\mathcal{P}_{L}=60 \ \mu$W in order to minimize saturation effects of the optical transition. The solid line is data fitting with Eq.~(1) leading to (a) $\theta=30\pm2^{\circ}$, (b) $\theta=73\pm5^{\circ}$, (c) $\theta=126\pm2^{\circ}$ and (d) $\theta=127\pm2^{\circ}$, in fair agreement with theoretical expectations. The slight discrepancy could be explained by a small tilt angle of the sample holder and by distortions of the laser polarization profile in the focal plane of the high-numerical-aperture microscope objective~\cite{Coolen2014}.}
\label{Fig1}
\end{figure*} 
\indent In most diamond samples, NV defects are oriented with equal probability along the four equivalent $\langle111\rangle$ crystal directions. Such a statistical distribution leads to a decreased sensitivity of sensing protocols relying on ensemble of NV defects, since only a quarter of NV spins are efficiently contributing to the detected signal~\cite{Rondin_2014,PhamPRB2013}. In the last years, several experiments have shown that chemical vapor deposition (CVD) techniques can be used to control the orientation of NV defects spontaneously created during the diamond growth. It was demonstrated that the efficiency of preferential alignment is closely tied to the CVD growth direction, and can thus be controlled by changing the orientation of the diamond substrate~\cite{PhamPRB2013,NewtonPRB2012,AtumiPRB2013,Lesik2014,Michl2014,Fukui2014,Ozawa2017,LesikDRM2015}. Preferential NV alignment was first observed in diamond crystals grown by CVD on $(110)$-oriented substrates~\cite{NewtonPRB2012}. In such samples, NV defects were mostly found along the $[111]$ and $[\bar{1}\bar{1}1]$ crystallographic axes, {\it i.e.} pointing out of the growth plane. These results were supported by quantum-mechanical simulations suggesting that the NV defect energy formation is minimized for orientations closest to the CVD growth direction, while in-plane NV orientations are highly improbable~\cite{AtumiPRB2013}. This explains why preferential NV alignment is not observed in conventional diamond samples grown by CVD on $(100)$-oriented substrates, because the angle between the NV axis and the growth direction is identical for the four NV defect orientations. These predictions were further confirmed by the observation of a perfect alignment of NV defects along the $[111]$ direction in diamond samples grown on (111)-oriented substrates~\cite{Lesik2014,Michl2014,Fukui2014,Ozawa2017}. Such a geometry seems ideal because it also leads to an optimized collection efficiency of the NV defect emission, potentially resulting in more sensitive quantum sensors~\cite{Appel2014}. However, $(111)$-oriented CVD growth is often plagued by the formation of penetration twins and extended defects resulting in diamond films with low crystalline quality~\cite{Kasu2003,Butler2008}. \\
\indent A promising alternative is offered by CVD growth on (113) substrates, which provides high quality crystals while preserving a significant alignment of NV defects along the $[111]$ direction~\cite{LesikDRM2015}. In this work, we demonstrate that NV orientation can be further controlled in $(113)$-oriented diamond samples by tuning the CVD growth temperature. More precisely, we show that preferential alignment along the $[111]$ direction is significantly improved when growth is performed at a lower temperature. This effect is then combined with temperature-dependent incorporation of NV defects during CVD growth~\cite{LesikDRM2015bis} to engineer thin diamond layers doped with aligned NV defects. These results illustrate that growth temperature can be exploited as a new degree of freedom to design optimized diamond samples for quantum sensing applications.

\indent The diamond crystals used in this work were synthesized by microwave plasma-assisted CVD using (113)-oriented optical grade diamond substrates provided by {\it Element 6}. In order to remove surface dislocations commonly induced by polishing, the substrates were first submitted to a H$_2$/O$_2$ (98/2) plasma etching directly inside the CVD reactor using a microwave power of $3$~kW and a pressure of $200$~mbar~\cite{Naamoun2012}. Diamond growth was then performed under high power density conditions~\cite{Achard2007} with a H$_2$/CH$_4$ (96/4) gas mixture at a total flow rate of $500$~sccm, resulting in a growth rate of about $10 \ \mu{\rm m/h}$. The growth temperature was modified by tuning the microwave power and the pressure in the CVD chamber. This procedure was facilitated by the fact that (113) CVD growth allows obtaining high quality and smooth films over a wide temperature range. Sample A was grown at $1000^{\circ}$C by using a microwave power of $3.5$~kW and a pressure of $250$~mbar, while sample B was grown at $800^{\circ}$C by decreasing both the microwave power and the pressure to $2.8$~kW and $180$~mbar, respectively. The growth temperature was measured with a 2-color pyrometer (Williamson Pro 82-40) focused on the diamond surface through a viewport. This method provides an accuracy of about $1^{\circ}$C for relative temperature variations during a given run of diamond growth. However due to the small dimensions of the crystal and the high plasma emissivity, getting an absolute temperature measurement remains a difficult task and the run-to-run temperature accuracy is estimated to about $\pm 30^{\circ}$C. 

\indent Individual NV defects were optically isolated in these (113)-oriented diamond samples with a scanning confocal microscope operating under green laser excitation at ambient conditions. The NV orientation distribution was inferred from electron spin resonance (ESR) spectra recorded for a large set of individual NV defects while applying a static magnetic field $\mathbf{B_0}$. In the limit of weak magnetic fields~\cite{Rondin_2014}, the ESR frequencies are given by $\nu_{\pm}=D\pm g\mu_{\rm B} |\mathbf{B_0}\cdot\mathbf{u_{\rm NV}}(\theta,\phi)|$, where $D\approx 2.87$~GHz is the zero-field splitting, $g\mu_{\rm B}=28$~GHz/T and $\mathbf{u_{\rm NV}}(\theta,\phi)$ denotes the unit vector of the NV defect orientation, which is characterized by the spherical angles $(\theta,\phi)$ in the $(x,y,z)$ sample reference frame [Fig.~1, top panels]. Here, the magnetic field $\mathbf{B_0}$ was chosen so that all four NV defect orientations could be easily distinguished by their ESR spectrum, as illustrated in the middle panels of Fig.~1. Each of these spectra was unambiguously associated with a given NV defect orientation by measuring the photoluminescence (PL) intensity while rotating the polarization angle of the linearly-polarized excitation laser~\cite{Santori2007}. Optical transitions of the NV defect involve two independent orthogonal dipoles, which are lying in the plane perpendicular to $\mathbf{u_{\rm NV}}(\theta,\phi)$~\cite{Epstein2005}. The excitation rate of the NV defect is then linked to the projection of these dipoles along the linear laser polarization in the (113) diamond surface plane. Unbalanced excitation of the two dipoles leads to a polarization-dependent PL intensity whose modulation depends on the NV defect orientation. For a laser excitation power much smaller than the saturation power of the optical transition, the NV defect PL intensity can be expressed as~\cite{ZhengPhD}
\begin{equation}
\mathcal{I}=\mathcal{I}_{\rm max}\left[1-\cos^2(\alpha-\phi)\sin^2\theta\right] \ ,
\end{equation}
where $\alpha$ is the laser polarization angle in the $(x,y)$ plane and $\mathcal{I}_{\rm max}$ depends on the collection efficiency of the NV defect emission. A minimum of PL intensity $\mathcal{I}_{\rm min}$ is obtained for $\alpha=\phi \ [\pi]$, while the contrast of the modulation is linked to the NV defect polar angle through $\cos^{2}\theta=\mathcal{I}_{\rm min}/\mathcal{I}_{\rm max}$~\cite{ZhengPhD,Dolan2013}. Polarization-dependent PL intensity diagrams recorded for the four possible NV defect orientations in a (113)-oriented diamond sample are shown in the bottom panels of Fig.~1. For $[111]$-oriented NV defects, the polar angle is minimal ($\theta=29.5^{\circ}$), leading to the weakest amplitude of the PL modulation [Fig.~1(a)]. Conversely, NV defects with a $[\bar{1}\bar{1}1]$ orientation are almost lying in the diamond surface plane ($\theta=80^{\circ}$), resulting in a PL modulation with maximal contrast [Fig.~1(b)]. For the two other orientations $\{[\bar{1}1\bar{1}],[1\bar{1}\bar{1}]\}$ with $\theta=121.5^{\circ}$, similar polarization diagrams are observed with a moderate contrast and a global phase shift resulting from different $\phi$ angles [Figs.~1(c),(d)]. This simple method enables us to link each ESR spectrum to a given NV defect orientation, without requiring complex rotations of the applied magnetic field. 
\begin{figure}[t]
\includegraphics[width =8.6cm]{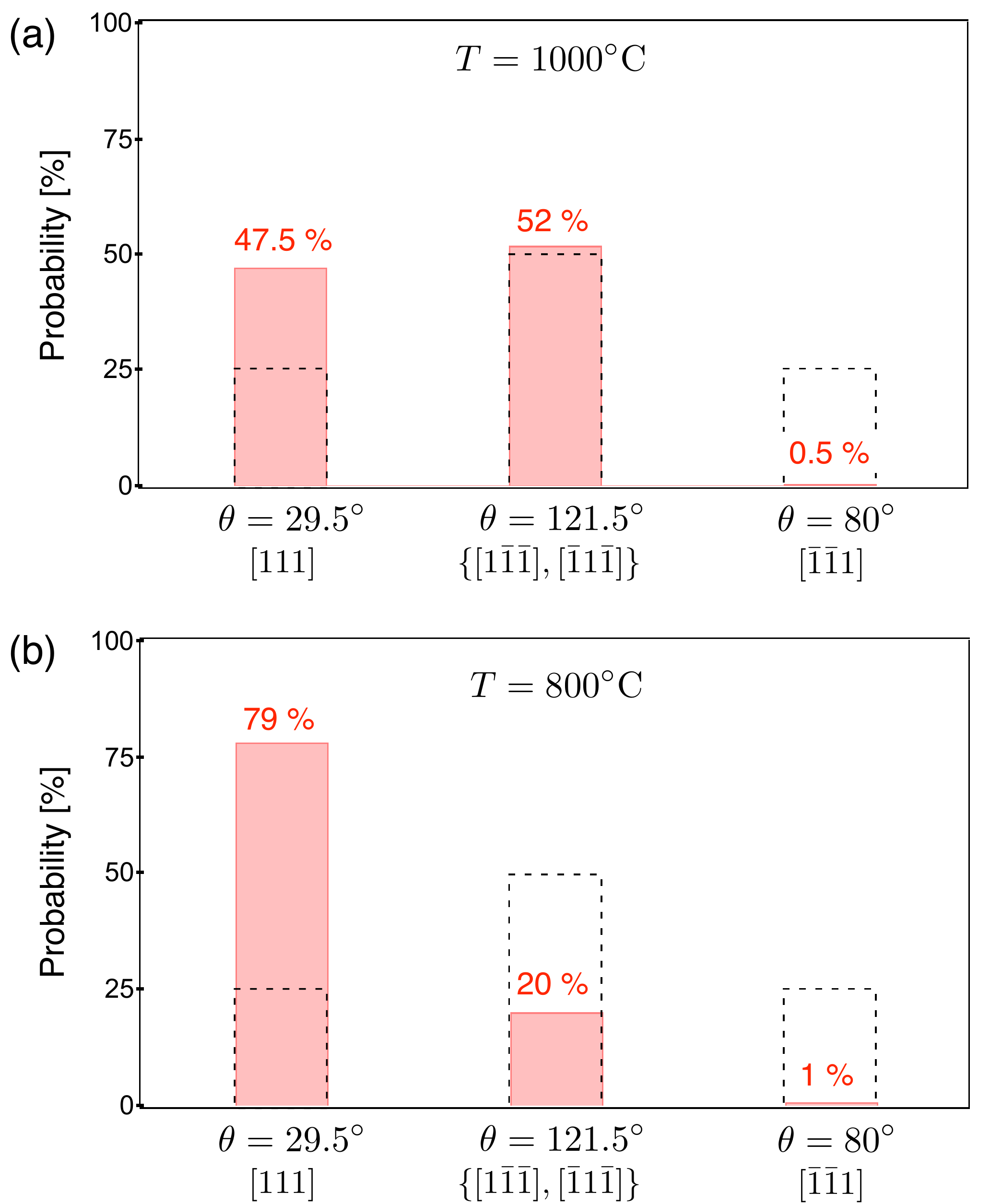}
\caption{Probability distribution of the NV defect orientation (a) in sample A and (b) in sample B. The black dotted bars indicate the expected distribution for a diamond sample without preferential alignment. These statistical distributions are obtained from orientation measurements over a set of about $200$ single NV defects.}
\end{figure}

The occurrence probability of each orientation was inferred from measurements over a large set ($\sim 200$) of individual NV defects. The results sorted by polar angle $\theta$ for two samples grown at different temperatures are displayed on Figure 2. In sample A grown at $T=1000^{\circ}$C, $[111]$-oriented defects are observed with a probability of about $50\%$, revealing a moderate effect of preferential orientation [Fig.~2(a)]. Interestingly, defects oriented along the $[\bar{1}\bar{1}1]$ direction, {\it i.e.} almost lying in the (113) diamond surface plane, are hardly observed. These results are in agreement with previous studies showing that NV defects are predominantly aligned along the orientation closest to the growth direction, while in-plane orientations are energetically unfavorable~\cite{PhamPRB2013,NewtonPRB2012,AtumiPRB2013,Lesik2014,Michl2014,Fukui2014,Ozawa2017,LesikDRM2015}. In order to analyze the impact of growth temperature on the NV orientation distribution, similar experiments were performed on sample B grown at $T=800^{\circ}$C. As illustrated in Fig.~2(b), decreasing the growth temperature significantly enhances the degree of preferential orientation along the $[111]$ direction that reaches $\sim 80\%$. Using a simple thermodynamic model, the statistical ratio of $[111]$ to $\{[\bar{1}1\bar{1}],[1\bar{1}\bar{1}]\}$ orientations is given by $\exp(-\Delta E/k_{\rm B}T)$ where $\Delta E$ is the difference of formation energy and $k_{\rm B}$ is the Boltzmann constant. Our experiments would suggest a difference of formation energy in the range of $100$~meV. This rough estimate might be useful for theoretical models aiming at describing the mechanisms of NV creation in CVD-grown diamond crystals~\cite{AtumiPRB2013}. We note that a previous study realized in a (113)-oriented diamond sample grown at $T=900^{\circ}$C has shown a preferential alignment of about $70\%$ along the $[111]$ direction~\cite{LesikDRM2015}, in agreement with the general trend of our experimental findings. \\
\begin{figure*}[t]
\includegraphics[width = 18cm]{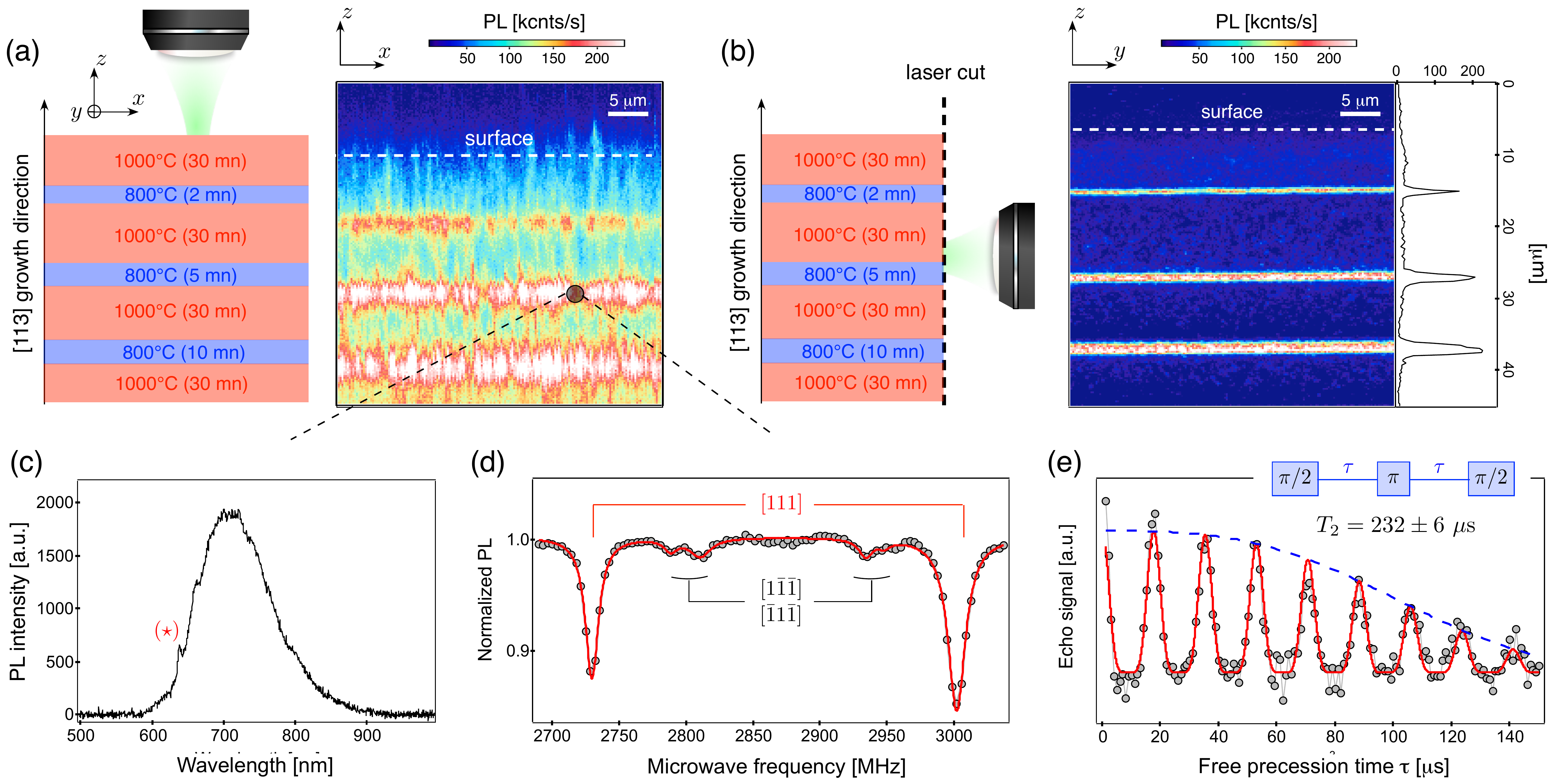}
\caption{(a) (left) Sketch of sample C grown at different temperatures and deposition times on a (113)-oriented diamond substrate. (right) Confocal PL raster scan recorded along the sample depth. The highest PL signal is recorded from the layers grown at the lowest temperature. (b) Confocal PL raster scan of a cross-section of sample C obtained by laser cutting. (c) PL spectrum recorded from a NV-doped layer. The symbol $(\star)$ indicates the characteristic zero-phonon line of the negatively-charged NV defect ($637$~nm). (d) Corresponding ESR spectrum measured while applying a static magnetic field which separates all NV defect orientations. (e) Spin-echo signal recorded for the subset of $[111]$-oriented NV defects with a static magnetic field of $\sim 50$~G applied along the $[111]$ axis. The inset shows the spin-echo sequence consisting of resonant microwave ($\pi/2$) and ($\pi$) pulses separated by a variable free precession time $\tau$.
}
\end{figure*}

Besides controlling the NV defect orientation, variations of the CVD growth temperature can also be employed to tune the NV defect density in diamond layers. Indeed, it was recently shown that the efficiency of NV defect creation is increased when the growth temperature is reduced~\cite{LesikDRM2015bis}. In the following, we make use of abrupt temperature variations during the CVD growth to obtain preferential alignment over {\it ensembles} of NV defects spatially localized in thin diamond layers, a key requirement for many NV-based quantum sensing applications. To this end, sample C was synthesized by alternating the growth temperature through quick modifications of the microwave power and the pressure in the CVD chamber. As sketched in Fig.~3(a), low temperature diamond layers grown at $800^{\circ}$C with an increasingly short deposition time were separated by high temperature layers grown at $1000^{\circ}$C. The typical time required to achieve such a temperature variation was around one minute. A slight amount of N$_2$ ($\sim 2$~ppm) was intentionally added into the gas phase throughout the entire growth run in order to increase the NV density. 

Figure~3(a) shows a confocal PL scan recorded along the depth of sample~C. The largest PL signal is detected from the diamond layers grown at the lowest temperature, illustrating the expected temperature-dependent creation of NV defects during the CVD growth. We note that NVs  are mainly created in the negatively-charged state in the doped layers. This is illustrated by the PL spectrum shown in Fig.~3(c), which do not reveal the characteristic emission of the neutral charge state NV$^0$ with a zero-phonon line at $575$~nm. Such an efficient charge-state stabilization is assigned to a large incorporation of substitutional nitrogen atoms N$_{\rm s}$, providing a n-type doping of the diamond crystal.\\
\indent Confocal PL images recorded from the sample top surface can hardly be used to estimate the thickness of NV-doped layers owing to the poor spatial resolution of the confocal microscope along the longitudinal direction [Fig.~3(a)]. In order to infer more precisely the spatial profile of doped layers, a cross-section of the sample was obtained through laser cutting and polishing. As shown in Fig.~3(b), a PL raster scan of the cross-section plane clearly reveals well-defined NV-doped layers with a spatial resolution limited by the lateral resolution of the confocal microscope ($\sim 350$~nm). The thinnest layer is obtained for the shortest deposition time (2 mn), leading to a PL profile with a linewidth (FWHM) in the order of $\sim 450$~nm. Considering the spatial resolution of the microscope, it corresponds to a thickness in the range of $300$~nm. Layers with thicknesses down to few tens of nanometers could be obtained in future by decreasing the diamond growth rate, for example by limiting the amount of methane introduced in the gas phase or by reducing the pressure/microwave power set. This will be facilitated by the wide range of suitable growth conditions provided by the (113) orientation. The density of NV defects in the doped layers was roughly estimated by comparing the PL intensity to that measured for a single NV defect with a laser excitation power well below the saturation of the optical transition. We obtain NV densities in the range of $\sim 10^{15} \ {\rm cm}^{-3}$, which could be further increased by at least one order of magnitude by adding more N$_2$ into the gas phase. \\
\indent As a final experiment we analyze the spin properties of NV-doped layers. A typical ESR spectrum is shown in Fig.~3(d), confirming the strong preferential alignment of NV defects along the [111] direction. This result illustrates that (113)-growth under lower temperatures is particularly suitable for creating ensembles of aligned NV defects. The spin coherence time ($T_2$) of the NV ensemble was then measured by applying a Hahn echo sequence. As shown in Fig.~3(e), the spin echo signal exhibits characteristic collapses and revivals induced by the interaction with a bath of $^{13}$C nuclear spin~\cite{Childress2006}. The decay of the envelope leads to a coherence time $T_2=232\pm6 \ \mu$s, which is similar to the one commonly obtained in conventional (100) crystals with identical isotopic purity~\cite{Mizuochi}.\\

In summary, a high degree of engineering of NV defect orientation, density and spatial localization is here achieved by tuning the temperature of CVD diamond growth. By relying on (113) oriented crystals, we have first shown that preferential alignment of NV defects along the [111] axis is significantly enhanced when the growth temperature is decreased. The combination of this effect with temperature-dependent creation efficiency of NV defects allows for the formation of thin diamond layers doped with dense ensembles of aligned NV defects. Importantly, such an engineering is realized while preserving the outstanding spin coherence properties of NV defects. These results establish that temperature-dependent CVD growth on (113)-oriented substrates is a promising approach for the design of optimized diamond samples for quantum sensing applications. \\

\noindent {\it Acknowledgments -} We acknowledge M. Markham and A. Edmonds from Element 6 for providing the (113) substrates used in this experiment. This research has been supported by the European Union Seventh Framework Program (FP7/2007-2013) under the project {\sc Diadems} and by the European Research Council  (ERC-StG-2014, {\sc Imagine}).


\begin{thebibliography}{}

\bibitem{DefectReview}
W. F. Koehl, H. Seo, G. Galli, and D. D. Awschalom, Designing defect spins for wafer-scale quantum technologies. {\it MRS Bulletin} {\bf 40}, 1146 (2015)

\bibitem{Dutt2007}
M. V. G. Dutt {\it et al.}, Quantum register based on individual electronic and nuclear spin qubits in diamond. {\it Science} {\bf 316}, 1312-1316 (2007)

\bibitem{Waldherr2014}
G. Waldherr {\it et al.}, Quantum error correction in a solid-state hybrid spin register, {\it Nature} {\bf 506}, 204-207 (2014)
    
\bibitem{Hensen2015}
B. Hensen {\it et al.}, Loophole-free Bell inequality violation using electron spins separated by 1.3 kilometres. {\it Nature} {\bf 356}, 682-686 (2015)

\bibitem{Kalb2017}
N. Kalb, A. A. Reiserer, P. C. Humphreys, J. J. W. Bakermans, S. J. Kamerling, N. H. Nickerson, S. C. Benjamin, D. J. Twitchen, M. Markham, and R. Hanson, Entanglement distillation between solid-state quantum network nodes. {\it Science} {\bf 356}, 928 (2017)

\bibitem{Neumann2013}
P. Neumann {\it et al.}, High-precision nanoscale temperature sensing using single defects in diamond. {\it Nano Lett.} {\bf 13}, 2738 (2013).

\bibitem{Toyli2013}
D. M. Toyli, C. F. de las Casas, D. J. Christle, V. V. Dobrovitski, and D. D. Awschalom, Fluorescence thermometry enhanced by the quantum coherence of single spins in diamond. {\it Proc. Natl Acad. Sci.} {\bf 110}, 8417 (2013).

\bibitem{Kucsko2013}
G. Kucsko, P. C. Maurer, N. Y. Yao, M. Kubo, H. J. Noh, P. K. Lo, H. Park, and M. D. Lukin, Nanometre-scale thermometry in a living cell. {\it Nature} {\bf 500}, 54 (2013).

\bibitem{DoldeNatPhys}
F. Dolde {\it et al.}, Electric-field sensing using single diamond spins. {\it Nat. Phys.} {\bf 7}, 459-463 (2011)

\bibitem{Maze}
J. R. Maze {\it et al.}, Nanoscale magnetic sensing with an individual electronic spin in diamond. {\it Nature} {\bf 455}, 644-647 (2008).

\bibitem{Gopi}
G. Balasubramanian {\it et al.}, Nanoscale imaging magnetometry with diamond spins under ambient conditions. {\it Nature} {\bf 455}, 648-651 (2008).

 \bibitem{Rondin_2014}
L. Rondin, J.-P. Tetienne, T. Hingant, J.-F. Roch, P. Maletinsky, and V. Jacques, Magnetometry with nitrogen-vacancy defects in diamond. {\it Rep. Prog. Phys.} {\bf 77}, 056503 (2014).

\bibitem{Acosta2009}
V. M. Acosta {\it et al.}, Diamonds with a high density of nitrogen-vacancy centers for magnetometry applications. {\it Phys. Rev. B} {\bf 80}, 115202 (2009)

\bibitem{Balasu2009}
G. Balasubramanian {\it et al.}, Ultralong spin coherence time in isotopically engineered diamond. {\it Nat. Mater.} {\bf 8}, 383 (2009)

\bibitem{Mizuochi}
N. Mizuochi {\it et al.}, Coherence of single spins coupled to a nuclear spin bath of varying density. {\it Phys. Rev. B} {\bf 80}, 041201(R) (2009)

\bibitem{Grotz2012}
B. Grotz {\it et al.}, Charge state manipulation of qubits in diamond. {\it Nat. Comm.} {\bf 3}, 729 (2012)

\bibitem{Pfender2017}
M. Pfender {\it et al.}, Protecting a Diamond Quantum Memory by Charge State Control. {\it Nano Lett.} {\bf 17}, 5931-5937 (2017)

\bibitem{DegenPRB2012}
B. K. Ofori-Okai, S. Pezzagna, K. Chang, M. Loretz, R. Schirhagl, Y. Tao, B. A. Moores, K. Groot-Berning, J. Meijer, and C. L. Degen, Spin properties of very shallow nitrogen vacancy defects in diamond. {\it Phys. Rev. B} {\bf 86}, 081406(R) (2012). 

\bibitem{TetiennePRB2018}
J.-P. Tetienne {\it et al.}, Spin properties of dense near-surface ensembles of nitrogen-vacancy centers in diamond. {\it Phys. Rev. B} {\bf 97}, 085402 (2018) 

 \bibitem{PhamPRB2013}
L. M. Pham, N. Bar-Gill, D. Le Sage, C. Belthangady, A. Stacey, M. Markham, D.J. Twitchen, M.D. Lukin, R.L. Walsworth, Enhanced metrology using preferential orientation of nitrogen-vacancy centers in diamond. {\it Phys. Rev. B} {\bf 86}, 121202 (2013).

\bibitem{NewtonPRB2012}
 A. M. Edmonds, U. F. S. D'Haenens-Johansson, R. J. Cruddace, M. E. Newton, K. M. C. Fu, C. Santori, R. G. Beausoleil, D. J. Twitchen, M. L. Markham, Production of oriented nitrogen-vacancy color centers in synthetic diamond. {\it Phys. Rev. B} {\bf 86}, 035201 (2012).
 
\bibitem{AtumiPRB2013}
M. K. Atumi, J. P. Goss, P. R. Briddon, and M. J. Rayson, Atomistic modeling of the polarization of nitrogen centers in diamond due to growth surface orientation. {\it Phys. Rev. B} {\bf 88}, 245301 (2013).

 \bibitem{Lesik2014}
M. Lesik, J. P. Tetienne, A. Tallaire, J. Achard, V. Mille, A. Gicquel, J.-F. Roch, and V. Jacques, Perfect preferential orientation of nitrogen-vacancy defects in a synthetic diamond sample. {\it Appl. Phys. Lett.} {\bf 104}, 113107 (2014).

 \bibitem{Michl2014}
 J. Michl {\it et al.}, Perfect alignment and preferential orientation of nitrogen-vacancy centers during chemical vapor deposition diamond growth on (111) surfaces. {\it Appl. Phys. Lett.} {\bf 104}, 102407 (2014).

 \bibitem{Fukui2014}
 T. Fukui {\it et al.}, Perfect selective alignment of nitrogen-vacancy centers in diamond. {\it Appl. Phys. Exp.} {\bf 7}, 055201 (2014).

 \bibitem{Ozawa2017}
H. Ozawa, K. Tahara, H. Ishiwata, M. Hatano, and T. Iwasaki, Formation of perfectly aligned nitrogen-vacancy-center ensembles in chemical-vapor-deposition-grown diamond (111). {\it Appl. Phys. Exp.} {\bf 10}, 045501 (2017).
 
 \bibitem{LesikDRM2015}
M. Lesik {\it et al.}, Preferential orientation of NV defects in CVD diamond films grown on (113)-oriented substrates. {\it Diam. Relat. Mater.} {\bf 56}, 47-53 (2015)

 \bibitem{Appel2014}
E. Neu, P. Appel, M. Ganzhorn, J. Miguel-Sanchez, M. Lesik, V. Mille, V. Jacques, A. Tallaire, J. Achard, and P. Maletinsky, Photonic nano-structures on (111)-oriented diamond. {\it Appl. Phys. Lett.} {\bf 104}, 153108 (2014)

\bibitem{Kasu2003}
M. Kasu and T. Makimoto, Formation of stacking faults containing microtwins in (111) chemical-vapor-deposited diamond homoepitaxial layers. {\it Appl. Phys. Lett.} {\bf 83}, 3465 (2003)

 \bibitem{Butler2008}
 J. Butler and I. Oleynik, A mechanism for crystal twinning in the growth of diamond by chemical vapour deposition. {\it Phil. Trans. R. Soc. A} {\bf 366}, 295-311 (2008).
 
 \bibitem{LesikDRM2015bis}
A .Tallaire {\it et al.}, Temperature dependent creation of nitrogen-vacancy centers in single crystal CVD diamond layers. {\it Diam. Relat. Mater.} {\bf 51}, 55-60 (2015)

\bibitem{Naamoun2012}
M. Naamoun, A. Tallaire, F. Silva, J. Achard, D. Doppelt and A. Gicquel, Etch-pit formation mechanism induced on HPHT and CVD diamond single crystals by H$_2$/O$_2$ plasma etching treatment, {\it Phys. Status Solidi A} {\bf 209}, 1715 (2012).

\bibitem{Achard2007}
J. Achard, F. Silva, A. Tallaire, X. Bonnin, G. Lombardi, K. Hassouni, and A. Gicquel, High quality MPACVD diamond single crystal growth: high microwave power density regime. {\it J. Phys. D} {\bf 40}, 6175-6188 (2007).

\bibitem{Santori2007}
T. P. Mayer Alegre, C. Santori, G. Medeiros-Ribeiro, and R. G. Beausoleil, Polarization-selective excitation of nitrogen vacancy centers in diamond. {\it Phys. Rev. B} {\bf 76}, 165205 (2007).

\bibitem{Epstein2005} 
R. J. Epstein, F. M. Mendoza, Y. K. Kato, and D. D. Awschalom, Anisotropic interactions of a single spin and dark-spin spectroscopy in diamond. {\it Nature Phys.} \textbf{1}, 94-98 (2005).

\bibitem{ZhengPhD}
D. Zheng, Study and manipulation of photoluminescent NV color center in diamond. Ph.D. Thesis, Ecole Normale Superieure de Cachan, 2010.

\bibitem{Dolan2013}
P. R. Dolan, X. Li, J. Storteboom, and M. Gu, Complete determination of the orientation of NV centers with radially polarized beams. {\it Opt. Exp.} {\bf 22}, 4379-4387 (2014).


\bibitem{Coolen2014}
C. Lethiec, J. Laverdant, H. Vallon, C. Javaux, B. Dubertret, J.-M. Frigerio, C. Schwob, L. Coolen, and A. Ma\^{\i}tre, Measurement of Three-Dimensional Dipole Orientation of a Single Fluorescent Nanoemitter by Emission Polarization Analysis. {\it Phys. Rev. X} {\bf 4}, 021037 (2014).

\bibitem{Childress2006}
L. Childress, M. V. Gurudev Dutt, J. M. Taylor, A. S. Zibrov, F. Jelezko, J. Wrachtrup, P. R. Hemmer, and M. D. Lukin, Coherent dynamics of coupled electron and nuclear spin qubits in diamond. {\it Science} {\bf 314}, 281-285 (2006).

\end{thebibliography}
\end{document}